\documentclass[twocolumn,pra,showpacs,superscriptaddress,amssymb,amsmath,amsmath,floatfix]{revtex4-1}
\usepackage{graphicx}
\usepackage{epstopdf}
\usepackage{bm}
\usepackage[colorlinks=true,linkcolor=blue,urlcolor=blue,citecolor=blue]{hyperref}
\usepackage{comment}
\usepackage{float}
\usepackage{cancel}
\usepackage{times}
\usepackage{amsmath, amsthm, amssymb, amsfonts}
\usepackage{xcolor}

\newcommand{\dd}{\mathrm{d}}

\newcommand{\ef}{\epsilon_{\mathrm{F}}}
\newcommand{\mv}{\langle V \rangle}

\newcommand{\affFUW}{Faculty of Physics, University of Warsaw, Pasteura 5, 02-093 Warsaw, Poland}

\begin{document}
	\title{The long-range interacting Fermi polaron}
	\author{Krzysztof Myśliwy and Krzysztof Jachymski}
	\affiliation{\affFUW}
    \date{\today}

 \begin{abstract}
We construct the simplest density functional for the problem of a single impurity interacting with a Fermi gas via a long--ranged potential using the Thomas--Fermi approach. We find that the Fermi polaron is fully bosonized in two dimensions, as the model results in a suitable Landau--Pekar functional known from the Bose polaron problem which describes a self--interacting impurity. In other dimensions, the impurity self--interacts with an infinite number of its own images, and no bosonization occurs. We discuss applications of our theory for the $2d$ exciton--polaron and the ionic polaron problem and compute the effective mass for these cases,  finding a self--trapping transition with order depending on the dimensionality. 
 \end{abstract}

\maketitle
{\it Introduction.}
Polarons are quasiparticles that emerge from dressing of impurities with the excitations of the quantum many-body medium that they are embedded in. Their introduction was motivated mainly by modeling the motion of electrons in solids~\cite{Pekar1946,Landau1948,Lee1953,Frohlich1954}. However, the idea of dressed states with modified energy and effective mass is general and can be applied to other systems regardless of their nature~\cite{Gross1962}. Polarons are nowadays among the central concepts in understanding quantum transport~\cite{Devreese2009}. Experiments with ultracold atomic gases enabled the study of polarons in both bosonic and fermionic media, and provided the possibility of tuning the interaction strength via Feshbach resonances, as well as new detection schemes which greatly advanced the field~\cite{Massignan2014}.

One of the most intriguing aspects of the impurity problem is the calculation of effective mass, an important quantity characterizing the polaron along with its energy and quasiparticle residue. In the perturbative weak interaction limit the correction to bare impurity mass is expected to be small. However, for strong interactions, as well as for strongly correlated media, the situation becomes drastically different and the mass renormalization can be exponentially large and diverge with the coupling constant~\cite{Rosch1999}. Furthermore,  a self--localization effect occurs in some approaches where the effective mass becomes infinite at finite interaction strength~\cite{Gross1959,Cucchietti2006,Yakaboylu2018,Seiringer2021}. It is still under debate whether such transition can occur in realistic polaron models.

The simplest variational Ansatz for the polaron wave function would consist of the free particle dressed with a single medium excitation. However, the interactions need not be perturbative ~\cite{Schmidt2018}. Full description of the system has to take into account the possibility of forming few- and many-body bound states and losing the quasiparticle picture, making fully numerical treatment such as Monte Carlo methods necessary~\cite{Luis2022}. However, field theoretical and variational approaches still bring valuable insights into the nature of the system. One example is the impact of Efimov trimers and larger clusters arising if one considers multiple Bogoliubov excitations of the Bose-Einstein condensate interacting with the impurity~\cite{Levinsen2015,Christianen2022,Christianen2023}. Another interesting case is when the impurity-medium interaction is attractive and long-ranged, allowing for large density increase around the impurity. This can be realized using ions embedded in cold gases~\cite{Tomza-2019,Christensen-2022} and in the bosonic case results in formation of many-body bound states~\cite{Astrakharchik2021}. Similar phenomena could be expected for excitons in a Fermi sea of electrons, as the electron-exciton interaction has the same nature as the ion-atom one and exciton polarons have already been observed~\cite{Verzelen2002,sidler2017fermi,Wang2018,Fey2020,Efimkin2021}.

While the polaron problem is clearly very challenging, there also exist remarkably simple theories of Fermi gases with large predictive power. In particular, Thomas-Fermi and density functional theories (DFT)~\cite{Lieb1973,Hohenberg1964} rely on effective single-particle description of the inhomogeneous medium and have proven to be extremely successful. Inspired by this approach, in this work we study the Fermi polaron problem in the regime where a DFT-like theory can be constructed, assuming the interaction potential is slowly varying. We demonstrate that large Fermi energy enables the series expansion of the energy functional, resulting in effective theory with a clear and intuitive structure. Remarkably, in two dimensions the bosonization occurs and one arrives at an analog of the seminal bosonic Pekar functional. 

{\it The model}. We consider a system of $N$ free fermions of mass $m_F$ and spin $s$ and one impurity particle of mass $m_I$ in $d$ dimensions enclosed in a large box of volume $L^d$. The impurity interacts with the fermions via a two-body potential $V$ which we assume to be bounded and to vanish sufficiently fast at infinity, but at the same time to be slowly varying, such that the system can be described using the Thomas-Fermi approximation with the functional
\begin{equation}\label{funct}
	\begin{split}
\mathcal{H}(\psi, \rho)&=\frac{\hbar^2}{2m_I}\int |\nabla \psi(x)|^2 \dd x+\frac{d\ef\rho^{-\frac{2}{d}}}{d+2} \int \rho(x)^{1+\frac{2}{d}} \dd x\\ &+\iint \rho(x) V(x-y)|\psi(y)|^2 \dd x \dd y
	\end{split}
\end{equation}
to be minimized with the normalization constraints $\int \rho(x) \dd x =N$ and $\qquad \int |\psi(x)|^2\dd x=1$. 
In the above, $\rho=N L^{-d}$ is the density of the uniform gas, and $\ef=\left(\frac{\Gamma(\frac{d+2}{2})}{2s+1}\right)^{\frac{2}{d}}\frac{2\pi \hbar^2}{m_F}\rho^{2/d}$ is the corresponding Fermi level, while $\psi$ is the wave function of the impurity, and $\rho(x)$ is the local density of fermions at point $x$. 
This approach appears natural in the context of long range forces: in fact, the system described by eq.~\eqref{funct} is essentially a collection of infinitesimally small boxes, each filled with the free Fermi gas with local density $\rho(x)$ having internal energy $\sim\rho(x)^{1+\frac{2}{d}}\dd x$ due to the Fermi pressure, and subject to an external potential with a source whose position is smeared out due to the quantum nature of the impurity, $\int |\psi(y)|^2 V(x-y)\dd y$.   

We first optimize the functional in eq.~\eqref{funct} over $\rho$ at fixed $\psi$. This immediately results in 
\begin{equation}\label{eq_dens}
\rho_{\psi}(x)=\rho\left(\frac{\mu-V_{\psi}(x)}{\ef}\right)^{\frac{d}{2}}\Theta(\mu-V_{\psi}(x))
\end{equation}
where
\begin{equation}
V_{\psi}(x)=(|\psi|^2\ast V) (x)= \int V(x-y)|\psi(y)|^2 \dd y, 
\end{equation}
 $\Theta$ denotes the Heaviside theta function, and $\mu$ is the chemical potential, satisfying
 \begin{equation}\label{chempot}
 \rho\int \left(\frac{\mu-V_{\psi}(x)}{\ef}\right)^{\frac{d}{2}}\Theta(\mu-V_{\psi}(x))dx=N.
\end{equation}
The theta functions can be removed, provided that the number of fermions, and hence the chemical potential, are large enough~\cite{SupMat}. The optimized $\rho_{\psi}$ can now be plugged back into~\eqref{funct}, which yields a non--linear functional of $\psi$ alone. In what follows, we shall perform this procedure in different dimensions separately. 

{\it Bosonization in two dimensions.}
The resulting equations are particularly simple for $d=2$. In fact, for fixed $N$, we can easily determine $\mu$ from~\eqref{eq_dens}
\begin{align}
\mu=\epsilon_{\mathrm{F}}+\frac{1}{L^2}\int V_{\psi}(x) \dd x=\epsilon_{\mathrm{F}}+ \langle V \rangle\, ,
\end{align} 
where $ \langle V \rangle$ denotes the mean value of the potential, which is independent of $\psi$ if the box is suitably large compared to the range of the potential, such that the integration might be extended to infinity.

Using the resulting density profile as well as  the expression for the Fermi level $\ef$ in two dimensions, the total energy can be written as
\begin{equation}
\mathcal{H}(\psi, \rho_{\psi})=E_{\mathrm{F}}\left(1+\frac{\mv}{\ef}\right)^2+\mathcal{E}^{\rm{Pek}}(\psi)
\end{equation}
where $E_{\mathrm{F}}=\frac{1}{2}N\ef$ is the ground state energy of the uniform Fermi gas, and the \emph{Pekar functional} $\mathcal{E}^{\rm{Pek}}(\psi)$  reads 
\begin{equation}\label{pek}
\begin{split}
\mathcal{E}^{\rm{Pek}}(\psi)=&\frac{\hbar^2}{2m}\int |\nabla \psi(x)|^2 \dd x\\&-\frac{g_s m_F}{4\pi\hbar^2} \iint |\psi(x)|^2 V^{(2)}(x-y)|\psi(y)|^2 \dd x \dd y.
\end{split}
\end{equation}
where we introduced the spin degeneracy factor $g_s=2s+1$ as well as the notation
\begin{equation}\label{conv}
V^{(2)}(x)=V\ast V(x)=\int V(x-y) V(y) \dd y\, .
\end{equation}
Eq.~\eqref{pek} is precisely the well--known Pekar functional describing the semiclassical theory of impurities interacting with \emph{bosonic} fields~\cite{Pekar1946}. Thus at the semiclassical level,  the Fermi polaron in 2d is essentially equivalent to the Bose polaron in the same dimension~\cite{SupMat}.

{\it Other dimensions -- the polaronic Droste effect.}
In dimension different than two, eq.~\eqref{eq_dens} for the chemical potential cannot be solved as easily. However, a perturbative expansion may be applied in this case in the regime when the local potential or its fluctuations are small, i.e. for $|V_{\psi}-\mv|\ll \ef$. Then $\mu\approx \ef$ and one can solve for $\mu$ and find the energy by expanding the density in $|V_{\psi}-\mv|/\ef$.
%Then we evaluate the energy~\eqref{funct} at the same order in $c_i/\ef$.  
We provide the details of this calculation in \cite{SupMat}.  The result is
\begin{equation}
\begin{split}
\mathcal{H}(\psi,\rho_{\psi})=&\frac{d}{d+2}N \ef \\ & +N \mv\left(1+\frac{d\mv}{4\ef}+\frac{(d-2)(d-4)\mv^2}{12\ef^2}\right)\\& +\mathcal{E}_{(3)}^{\rm{Pek}}(\psi)
\end{split}
\end{equation}
with the following generalized version of the Pekar functional:
\begin{align}\label{pek3}
\mathcal{E}_{(3)}^{\rm{Pek}}(\psi)=&\frac{\hbar^2}{2m}\int |\nabla \psi(x)|^2 \dd x \nonumber\\ & -\frac{d}{4}\frac{\rho}{\ef}\left(1+\frac{(d-2)(d-4)\langle V \rangle}{2d\ef}\right)W_2(\psi) \nonumber\\ &+\frac{(d-2)(d-4)}{24}\frac{\rho}{\ef^2}W_3(\psi)
\end{align} 
with the already encountered two--body interaction
\begin{equation}
W_2(\psi)=\int_{{\left(\mathbb{R}^d\right)}^{\times 2}} |\psi(x_1)|^2 V^{(2)}(x_1-x_2)|\psi(x_2)|^2 
\end{equation}
and a newly emerging \emph{three-body interaction}
\begin{equation}
W_3(\psi)=\int_{{\left(\mathbb{R}^d\right)}^{\times 3}} |\psi(x_1)|^2|\psi(x_2)|^2|\psi(x_3)|^2V^{(3)}(x_1,x_2,x_3)
\end{equation}
with the three--body potential 
\begin{equation}
V^{(3)}(x_1,x_2,x_3)=\int V(x_1-y)V(x_2-y)V(x_3-y) \dd y.
\end{equation}
Note that if one sets $d=2$, the results of the preceding section are recovered, although the latter are not perturbative. Moreover, in $d\neq2$, the density enters the functional explicitly as a parameter modifying the coupling, while in $d=2$ the functional does not depend on the density. The main observation is, however, that the three--body term appears which is not present in the $2d$ Pekar functional for fermions, or for bosons in any dimensionality. One can say that here the image of the impurity in the medium creates its own image in turn, and all three objects interact with each other, also via genuine three--body forces. Note that one can proceed with the expansion procedure as above to higher orders. The inclusion of further terms in the series in inverse powers of $\ef$ yields a functional with the general structure 
\begin{align}
\mathcal{E}_{(\infty)}^{\rm{Pek}}(\psi)=&\frac{\hbar^2}{2m}\int |\nabla \psi(x)|^2 \dd x\nonumber \\&+\sum_{k=1}^\infty \frac{\rho}{\ef^k} \int_{\mathbb{R}^{d(k+1)}} G^{(k)}(x_1,\cdots ,x_{k+1})\prod_{i=1}^{k+1}|\psi(x_i)|^2
\end{align} 
with appropriate $k+1$--body kernels $ G^{(k)}(x_1,\cdots ,x_{k+1})$ depending on $V$. In the pictorial language of Pekar's theory, the fermionic correlations lead to an emergence of infinite mirror images of the impurity imprinted in the Fermi medium, resembling the self--similar structures employed by artists. Borrowing from the fine arts language, we describe this phenomenon as the \emph{polaronic Droste effect}~\footnote{see Droste effect Wikipedia article, \url{https://en.wikipedia.org/wiki/Droste_effect}}.
%%%%%%%%%%%%%%%%%%%%%%%%%%

{\it Effective mass and localization transition.}
In the next step, we are going to study the predictions of the semiclassical theory of the Fermi polaron. Apart from the computation of the ground state energy of the polaron, an important quantity to estimate is its effective mass. We adopt an approach to the effective mass problem inspired by \cite{Lieb2014}. The idea is to place the impurity particle into a very shallow harmonic trap of low frequency $\omega$, such that the Thomas--Fermi functional reads now
\begin{equation}\label{functom}
	\begin{split}
		\mathcal{H}(\psi, \rho;\omega)&=\frac{\hbar^2}{2m_I}\int |\nabla \psi(x)|^2 \dd x+\frac{d\ef\rho^{-\frac{2}{d}}}{d+2} \int \rho(x)^{1+\frac{2}{d}} \dd x\\ &+\iint \rho(x) V(x-y)|\psi(y)|^2 \dd x \dd y\\&+\frac{m_I \omega^2}{2}\int x^2|\psi(x)|^2 \dd x.
	\end{split}
\end{equation}
If the polaron is indeed formed and the impurity and the gas behave as one entity with effective mass $M_{\rm{eff}}$, then the difference between the ground state energies of the functional \eqref{functom} at small $\omega$ and $\omega=0$ should be well described by the ground state energy of the hamiltonian $\frac{\hbar^2}{2M_{\rm{eff}}}(-i \nabla_x)^2+\frac{m_I\omega^2}{2}x^2$, namely $\frac{d}{2}\hbar \omega\sqrt{\frac{m_I}{M_{\rm{eff}}}}$.
Accordingly, we define the effective mass as      
\begin{equation}\label{mass-def}
	M_{\rm{eff}}=\lim_{\omega\rightarrow 0} \frac{d^2 m_I \hbar^2 \omega^2}{4(E^{\rm{TF}}(\omega)-E^{\rm{TF}}(0))^2}
\end{equation}
with $E^{\rm{TF}}(\omega)$ being the minimum energy of the functional \eqref{functom}, assuming the limit exists. In other words, we expect a shift in the frequency of oscillations of the impurity immersed in a Fermi gas as compared to the motion in vacuum, and attribute this shift to the mass renormalization. This directly corresponds to possible experimental effective mass measurements in ultracold atomic setups. In practice, we are going to estimate $E^{\rm{TF}}(\omega)-E^{\rm{TF}}(0)$ e.g. by variational methods, and fit the results to a parabolic curve for sufficiently small $\omega$. The limit \eqref{mass-def} can then be obtained from the value of the fitted parameter in front of the linear term. 
%The intuition behind our definition of the effective mass is depicted in Fig. \ref{massdefok}. 

\begin{figure}
	\centering
	\includegraphics[scale=1.0]{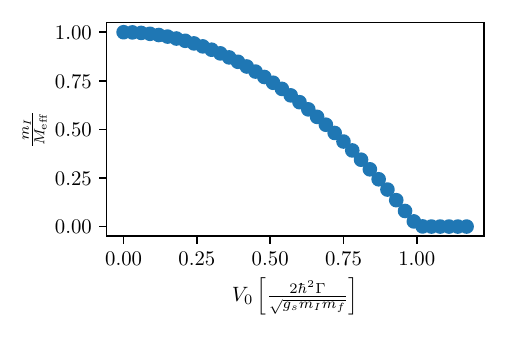}
	\caption{Two--dimensional inverse effective mass in units of the impurity mass as a function of the coupling strength in units of $2\hbar \Gamma/\sqrt{g_s m_I m_F}$. The mass diverges continuously at critical coupling, and thus the self-trapping transition is of \emph{second order}.}\label{m2dG}
\end{figure}

\begin{figure}
	\centering
	\includegraphics[scale=1.0]{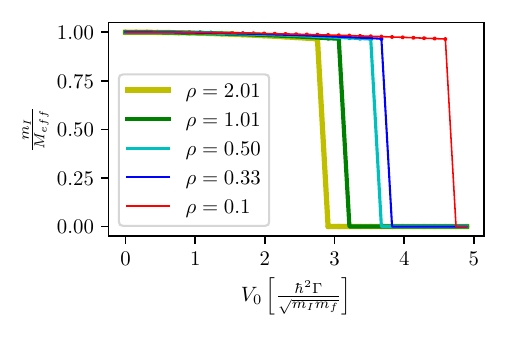}
	\caption{Three--dimensional inverse effective mass in units of the impurity mass, plotted against the coupling. The fermions are fully polarized such that $g_s=1$. The full lines present the results obtained from the Pekar functional with three--body terms included, the dots indicate truncation to two--body terms (no Droste effect). The densities are given in units of the volume set by the potential range $\Gamma^{3/2}$. As the effective mass jumps to infinity discontinuously at critical coupling, the trapping transition is of \emph{first order}.}\label{m3dG}
\end{figure}

As a first simple example, we choose a Gaussian--type potential 
\begin{equation}\label{gauss_pot}
	V(x)=-V_0 \exp\left(-\Gamma x^2\right)
\end{equation} 
with the coupling amplitude $V_0>0$ and the inverse range squared $\Gamma$ as the underlying impurity--fermion two--body potential, and estimating the ground state energies of the respective Pekar functionals in $d=2$ and $d=3$,~\eqref{pek} and~\eqref{pek3} respectively,  with a one--parameter trial wave function 
\begin{equation}\label{psib}
	\psi_u(x)=\left(\frac{2u}{\pi}\right)^{\frac{d}{4}} \exp\left(-u x^2\right), \quad u>0.
\end{equation}
This case can be largely treated analytically~\cite{SupMat}. The main conclusion here is the presence of a sharp \emph{localization transition}: depending on the value of the coupling $V_0$, the optimal $u$ is either zero, corresponding to an essentially free particle, or finite, which marks the onset of a bound state. The transition is of second order in $d=2$, with a continuous change in the value of the optimal $u$ from zero to positive values, and of first order in $d=3$ where the change is abrupt. In the "polaronic" phase where $u=0$ the effective mass is finite and lies always above the pure impurity mass, pointing at the quasi--particle formation in this regime, whereas when the bound state appears, the effective mass becomes infinite. In accordance with the order of the transition in the respective dimensions, the divergence of the effective mass as the coupling approaches the critical value necessary for binding is either continuous in $d=2$ or sharp in $d=3$. The results are summarized in Fig.~\ref{m2dG}-\ref{m3dG}. Note that in $3d$, the impact of the three--body terms on the effective mass in negligible in the polaronic phase even for moderate densities, as might be expected since in this regime $V_{\psi}\approx \langle V \rangle$ due to the large spread of the wave function,

Let us emphasize that self--trapping of the impurity is not due to the specific choice of the Gaussian potential or the trial wave function, but is a feature of the non--linearity of the Pekar functional, and its occurrence for sufficiently well-behaved potentials in $d=2$ and $d=3$ can indeed be proven with the use of the Sobolev inequality~\cite{Mys2023}. However, we do not expect such a transition to occur in a strict sense in real systems, and view it rather as an enhancement of the crossover between the polaron regime and the formation of cluster--like many-body bound state. %If so, then the crossover, predicted to happen in Fermi polarons with short range forces \cite{} as well as Bose polarons \cite{}, unifies all the instances of the polaron problem in cold gases.  

{\it Applications.}
We now apply the model to two simple and experimentally relevant cases in which the impurity--fermion pair consists of a charge and a neutral polarizable entity, such that the underlying two--body potential at long range comes from electrostatic induction and decays as $~r^{-4}$. We use regularized interaction with a finite depth
\begin{equation}\label{ionpol}
	V(r)=-\frac{C_4}{(r^2+b^2)^2}\, ,
\end{equation} 
where $b>0$ is the regularizing length scale, while $C_4=\frac{1}{2}q^2 \alpha$ with $q$ being the charge and $\alpha$ is the polarizability of the neutral object in question. This potential comes along with the length and energy scales $R^*=\sqrt{2m_rC_4/\hbar^2}$ and $E^*=\hbar^2/(2 m_r R^{*2})$, respectively, with $m_r$ denoting the reduced mass of the impurity--fermion pair. Using a Feshbach resonance it is possible to tune experimentally the scattering length of the potential, which we model by tuning the value of $b$.

\begin{figure}
	\centering
	\includegraphics[scale=1.0]{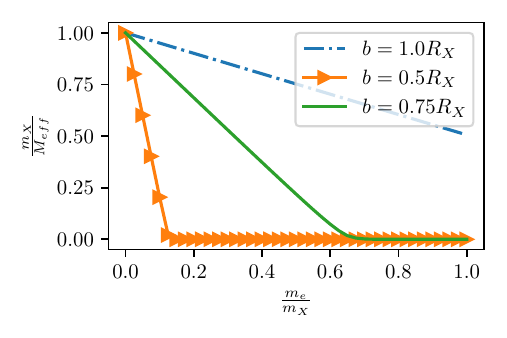}
	\caption{The inverse effective mass of the exciton polaron as a function of the electron to exciton mass ratio in the material, for different values of the length scale $b$. Here, $R_X=\sqrt{C_4 m_X}/\hbar^2$ with $m_X$ the exciton mass. }\label{mXpol}
\end{figure}

First we study the exciton--polaron problem in $2d$, i.e. a mobile exciton interacting with free electrons in a $2d$ layer, taking the functional~\eqref{pek} with the potential~\eqref{ionpol}, and using Gaussian trial functions. Assuming that the exciton polarizability is independent of its mass and neglecting any external potentials, the relevant parameter quantifying the coupling is given by the ratio of the effective masses of the electron and the exciton in the material. In Fig.~\ref{mXpol} we present the effective mass \iffalse and the energy and \fi of the exciton--polaron as a function of the electron to exciton mass ratio at different values of the regularizing scale $b$. As in the simple Gaussian model, we encounter a self--trapping transition of second order.

\begin{figure}
	\includegraphics[scale=1.0]{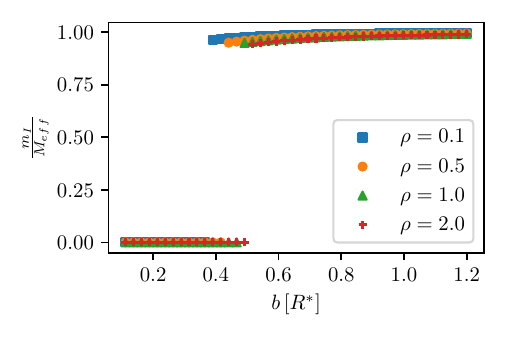}
	\includegraphics[scale=1.0]{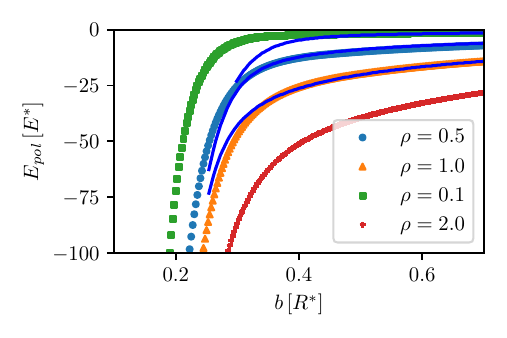}
	\caption{Upper: the inverse of the effective mass of the ionic polaron in the case where the impurity and the (polarized) fermions are of equal mass. Lower: the ground state energy of the polaron. The thin full lines are the results of the ladder approximation at the same densities \cite{Christensen-2022}, which agree for $b$ large as both theories converge to the mean--field limit there. Dimer formation in the two--body problem occurs at $b\sim 0.6R^*$ \cite{Christensen-2022}. The densities are given in units of $\left(R^*\right) ^{-3}$.}\label{m3dEqMass}
\end{figure}

Second, we use the $3d$ Pekar functional with the potential~\eqref{ionpol} for the ionic Fermi polaron problem, again with a Gaussian trial function and without the inclusion of many--body terms. As we are interested principally in the computation of the effective mass, the latter approximation should be enough for our purpose, even at moderate densities, see~Fig \ref{m3dG}.  We apply the theory for the case of equal masses as well as to the experimentally more interesting mass--imbalanced case of a $\mathrm{Ba}$ ion immersed in a cold gas of ${}^6\mathrm{Li}$ atoms~\cite{Weckesser2021}. The energy of the equal mass case is depicted in the lower panel of Fig.~\ref{m3dEqMass} and is compared to the result of Christensen {\it et al.}~\cite{Christensen-2022}, who calculated the polaron energy within the ladder approximation. We note that the two curves agree perfectly in the large $b$ limit where the potential is very shallow and mean field theory should be strict. For deeper potentials both approaches provide qualitatively similar results.
The effective mass follows the lines found in the Gaussian model and diverges disconitunously for values of $b$ corresponding to dimer formation in the two--body problem, cf. the upper panel of Fig~\ref{m3dEqMass}, which marks the presence of the self--trapping transition of first order.
We remark that recent Monte--Carlo calculations of the effective mass of the ionic polaron~\cite{Ardila2023} display an abrupt increase for values of $b$ in the same range, validating the use of the semiclassical theory.

\begin{figure}[t]
	\includegraphics[scale=1.0]{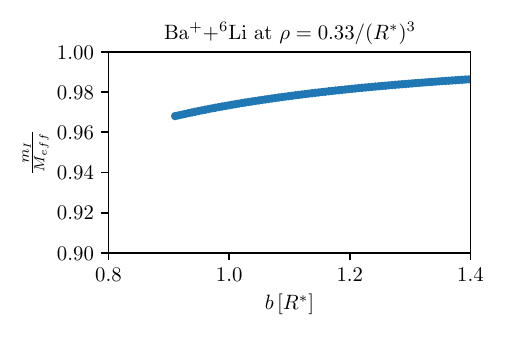}
	\includegraphics[scale=1.0]{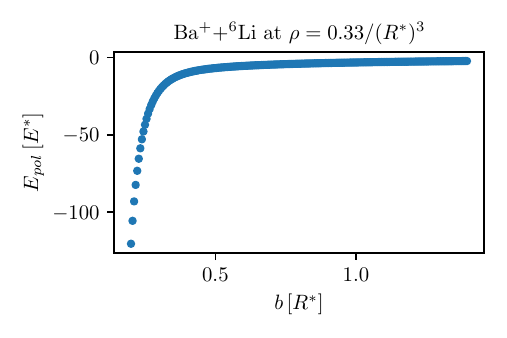}
	\caption{Upper: the inverse of the effective mass of the ionic polaron made of a $\rm{Ba}$ ion in a ${}^6\rm{Li}$ gas at density $\rho=10^{15} \rm{cm}^{-3}$. The line terminates at self--trapping where the inverse effective mass vanishes. Lower: the ground state energy of the ionic polaron for the same parameters.}\label{liba}
\end{figure}

The case of the ionic Fermi polaron with mass imbalance gives similar results, as depicted in Fig. \ref{liba} where we plot the effective mass as a function of $b$ for the experimentally relevant value of the density~$\rho=10^{15}$cm$^{-3}$. For this density, we predict a moderate mass increase of up to $3\%$ in the polaron regime $b\geq b_c\approx 0.9 R^*$ and its abrupt enhancement for $b\leq b_c$.  

{\it Summary.}
We have developed a simple theory of an impurity interacting with a free Fermi gas via a two--body long--range potential. Under these conditions, the local density approximation in the spirit of Thomas--Fermi can be applied to the description of the Fermi gas, which results in a suitable semiclassical energy functional depending on the fermion number density $\rho$ and the impurity wave function $\psi$. We found that in two dimensions this leads to the non--linear Pekar functional known from the bosonic counterpart of the problem,  describing a self--interacting impurity. This suggests that the (long--ranged) Bose and Fermi polarons in 2D may have similar physical properties not present in other dimensions. 
\\ \indent In dimensions other than two, a perturbative expansion can be applied in the quasi--free gas limit. It gives rise to a new Pekar functional with multiple self--interactions of the impurity with its own images, whose number increases with the order of the expansion. Inspired by terminology from the fine arts, we refer to this phenomenon as the \emph{polaronic Droste effect}. Nevertheless, for the experimentally relevant scenario of the ion-atom system the images do not significantly impact the value of the effective mass.
\\ \indent Interestingly, we found that a localization transition takes place in the system, as particle either forms a polaron with renormalized mass or an immobile many-body bound state. The presence of such transition marks a boosted version of the smooth crossover between polarons and many--body clusters which is likely to be found in the full quantum treatment and in the laboratory. 
\\ \indent Our approach can be viewed as a kind of the density functional theory for the Fermi polaron problem in its simplest form, which already provides insight into the physics involved. We expect our results to act as useful guidelines for future experiments and theories to follow, in particular the measurements of the effective mass in various dimensions and its dependence on the gas density and the impurity--fermion mass ratio. All these parameters can easily be varied within our approach, while the model itself remains open to further analysis.

\emph{Acknowledgements}. 
We thank R. Seiringer for helpful discussions on the properties of the Pekar functional in $2d$, L.P. Ardila for the discussions of his results in the equal mass case, and E. Christensen for sharing his results in the ladder approximation. This work was supported by the National Science Centre of Poland Grant 2020/37/B/ST2/00486 and the Polish National Agency for Academic Exchange (NAWA) via the Polish Returns 2019 programme.

\bibliography{refs}

\appendix

\section{Perturbation expansion}
We shall give a few more details about the perturbation expansion which leads to the polaronic Droste effect described by the Pekar functional with many--body terms. 
\\ 
\indent As pointed out in the main text, the regime we are interested in corresponds to situations where the chemical potential is close to $\ef$, which is its value for the free Fermi gas at zero temperature. We assume that the impurity--fermion potential $V$ is bounded, integrable and decays sufficiently fast such that all boundary effects are negligible, in particular integrals involving $V$ may be extended to infinity. For simplicity, we shift the interaction potential $V\rightarrow V-\langle V \rangle $ and work with a $V$ that has mean value zero. When considering the final result, one simply has to add a constant term $N\langle V \rangle$ to the energy functional. Further, in order to control the expansion, we substitute $V_{\psi}\rightarrow \lambda V_{\psi}$ where $\lambda>0$ is meant to be a small characteristic energy scale resulting either from the weakness of the potential $V$ or the large spread of the wave function $\psi$, in which case $V_{\psi}$ is indeed small as we assume that $V$ has mean value zero. Our expansion is then in the dimensionless parameter $\lambda/\ef$, and at the end $\lambda$ is set equal to unity. Accordingly, we write 
\begin{equation}
	\mu=\ef\left(1+\frac{\lambda}{\ef} \mu_0+\left(\frac{\lambda}{\ef}\right)^2 \mu_1 +\left(\frac{\lambda}{\ef}\right)^3 \mu_2+o((\lambda/\ef)^4)\right)
\end{equation}
and plug this into the formula for the optimal density profile given $\psi$, i.e., 
\begin{equation}
	\rho_{\psi}(x)=\rho\left(\frac{\mu-\lambda  V_{\psi}}{\ef}\right)^{\frac{d}{2}}
\end{equation}
(the Heaviside function can be dropped since in the assumed regime of validity of the expansion it clearly holds that $\mu>\lambda V_{\psi}(x)$ for all $x$). The application of the generalized binomial expansion 
\begin{equation}\label{binom_app}
	(1+x)^{\alpha}=\sum_{k=0}^{\infty}\frac{\Gamma(\alpha+1)}{\Gamma(\alpha+1-k)k!}x^k\equiv\sum_{k=0}^{\infty} {\alpha\choose k }x^k
\end{equation} with $\Gamma(\cdot)$ denoting the Euler Gamma function yields 
\begin{align*}
	\rho_{\psi}(x)=&\rho\left(1+\frac{\lambda}{\ef}D(\mu_0-V_{\psi}(x))\right.\\
	&+\frac{\lambda}{\ef}^2\left(D\mu_1+{D\choose 2}(\mu_0-V_{\psi}(x))^2\right)+\\
	-	&\left(\frac{\lambda}{\ef}\right)^3\left(2{D \choose 2}\mu_1(\mu_0-V_{\psi}(x))+D\mu_2\right. + \\ 
	&\left.\left. {D\choose 3}(\mu_0-V_{\psi}(x))^3 \right)+o((\lambda/\ef)^{4})\right)
	\end{align*}
where we introduced $D=\frac{d}{2}$ for the sake of transparency. Then $\int \rho_{\psi}(x) \dd x-N$ is a polynomial in $\lambda$. Since $\int \rho_{\psi}(x)\dd x =N$, all coefficients of this polynomial have to vanish, which leads to 
\begin{align*}
	&\mu_0=0 \\
	&\mu_1=-\frac{1}{D}{D\choose 2} \frac{\int V_{\psi}(x)^2 \dd x}{L^d} \\
	&\mu_2=\frac{1}{D}{D \choose 3}  \frac{\int V_{\psi}(x)^3 \dd x}{L^d}.
\end{align*}
(Recall that $\int V_{\psi}(x)=\int V=0$ -- the first equality follows from the assumption that $V$ vanishes fast enough so that the integration might be extended to infinity, and the second since we have shifted the mean value from the potential at the beginning.) 
We plug these back into the density profile and compute resulting energy from the functional $\mathcal{H}(\rho,\psi)$. With the use of \eqref{binom_app}, the internal energy of the fermions is then evaluated to 
\begin{align*}
	&\frac{d}{d+2}\ef \rho^{-\frac{2}{d}}\int \rho(x)^{1+\frac{2}{d}}\dd x=\\ &\frac{d}{d+2}N\ef +\frac{\lambda^2}{\ef}\frac{d}{d+2}{D' \choose 2} D^2\int V_{\psi}(x)^2 \dd x\\ &-\frac{\lambda^3}{\ef^2}\left({D' \choose 3}D^3+2 {D \choose 2} {D'\choose 2} D\right)\int V_{\psi}(x)^3 \dd x \\&+o(\lambda^4\ef^{-3})
\end{align*} 
where $D'=1+\frac{2}{d}$. The interaction energy at the same order reads 
\begin{align*}
	&\lambda \int \rho(x) V_{\psi}(x)\dd x =-\frac{\lambda^2}{\ef}\rho D\int V_{\psi}(x)^2 \dd x\\&+\frac{\lambda^3}{\ef^2}\rho{D \choose 2} \int V_{\psi}(x)^3 \dd x+o(\lambda^4 \ef^{-3}).
\end{align*}
After simple manipulations, we arrive at the final expression
\begin{align}
	\mathcal{H}(\psi, \rho_{\psi})=&\frac{\hbar^2}{2m_I}\int |\nabla \psi(x)|^2 \dd x\nonumber \\ \nonumber &-\frac{d}{4}\frac{\rho}{\ef} \int \tilde{V}_{\psi}(x)^2 \dd x\\&+\frac{(d-2)(d-4)}{24}\frac{\rho}{\ef^2}\int \tilde{V}_{\psi}(x)^3 \dd x +N \langle V \rangle
\end{align} where we have put $\lambda=1$ and denoted explicitly the shifted potential $\tilde{V}_{\psi}=V_{\psi}-\langle V \rangle$. By noting that $\int V_{\psi}^2 \dd x= W_2(\psi), \int V_{\psi}^3 \dd x= W_3(\psi)$, it is straightforward to arrive at the form given in the main text. It is evident from the procedure sketched that at higher orders in $\lambda$, many-- body terms  will emerge from expressions like $\int V_{\psi}(x)^k \dd x$ which appear at higher orders in the binomial expansion of the energy.

\section{Self--trapping within the Gaussian Ansatz}
Now we turn to the analysis of the Pekar functional (given by Eq.~(7) in the main text) using the Gaussian--type potential which is commonly chosen as a model of finite--ranged interaction
\begin{equation}\label{gauss_pot}
	V(x)=-V_0 \exp\left(-\Gamma x^2\right)
\end{equation} 
with the coupling amplitude $V_0>0$ and the inverse range squared $\Gamma$. 

Then a simple calculation shows that
\begin{equation}
	V^{(2)}(x)=\frac{\pi V_0^2}{2\Gamma}\exp\left(-\frac{\Gamma}{2}x^2\right).
\end{equation}
As indicated in the main text, we estimate the ground state energy of $\mathcal{E}^{\rm{Pek}}$ using a one-parameter trial wave function 
\begin{equation}\label{psib}
	\psi_u(x)=\left(\frac{2u}{\pi}\right)^{d/4} \exp\left(-u x^2\right), \quad u>0.
\end{equation}
We start with $d=2$. 
The resulting energy is very easy to calculate, and reads 
\begin{equation}
	E(u)\equiv  \mathcal{E}^{\rm{Pek}}(\psi_u)=\frac{\hbar^2}{m_I}u-\frac{g_s m_FV_0^2}{8\Gamma \hbar^2} \frac{u}{u+\frac{\Gamma}{2}}.
\end{equation}
Already this simple approach leads to the occurrence of a localisation transition, i.e. the phenomenon that a bound state appears only for sufficiently strong couplings. In fact, the positive root of the derivative of $E(u)$ is given by 
\begin{equation}
	%b_{+}=\sqrt{\frac{\pi V_0^2m}{8a_0\hbar^2}}-\frac{\Gamma}{2}
	u_+=\frac{1}{2}\left(\frac{V_0}{2\hbar}\sqrt{g_s m_I m_F}-\Gamma\right)
\end{equation}
and appears if and only if 
\begin{equation}
	V_0\geq V_c=\frac{2\Gamma\hbar^2}{\sqrt {g_s m_F m_I }}.
\end{equation}
Otherwise, the lowest energy attainable by means of $\psi_u$ equals zero, and corresponds to the ground state of the free particle in the box. Thus, for $V_0<V_c$, the surrounding gas does not localise the particle, in contrast to the situation encountered in Pekar's theory of electrons in polar crystals. At $V_0>V_c$ binding does occur, and the particle gets trapped by the surrounding medium.
If we now again take \eqref{gauss_pot} as the particle--impurity potential and define also 
\begin{equation}
	E(u;\omega)=\mathcal{E}^{\rm{Pek}}(\psi_u;\omega); \quad E_0(\omega)=\inf_{u>0} E(u,\omega)
\end{equation} 
with $\psi_u$ denoting the class of functions \eqref{psib}, then the effective mass in the \emph{Gaussian} approach will be 
\begin{equation}\label{gaussmass}
	M^{\rm{G}}_{\rm{eff}}=\lim_{\omega\rightarrow 0} \frac{d^2 m_I \hbar^2 \omega^2}{4(E_0(\omega)-E_0(0))^2}.
\end{equation}
In $2d$,  it is then straightforward to find
\begin{equation}
	E(u;\omega)=\frac{\hbar^2}{m_I}u-\frac{g_s m_F V_0^2}{8\Gamma \hbar^2} \frac{u}{u+\frac{\Gamma}{2}}+\frac{m_I\omega^2}{4u}.
\end{equation} 
The condition for the zero of the derivative 
\begin{equation}
	\frac{\hbar^2}{m}-\frac{g_s m_F V_0^2}{16 \hbar^2}\frac{1}{(u+\frac{\Gamma}{2})^2}-\frac{m_I\omega^2}{4u^2}=0  
\end{equation}
leads to the quartic equation 
\begin{equation}\label{quartic}
	\frac{\hbar^2}{m}u^2\left(u+\frac{\Gamma}{2}\right)^2-\frac{g_s m_FV_0^2}{16 \hbar^2}u^2-\frac{m_I\omega^2}{4}\left(u+\frac{\Gamma}{2}\right)^2=0 .
\end{equation}
whose unperturbed ($\omega=0$) nonnegative solutions are $u_{p}=0$ in the unbound regime $(V_0<V_c)$ and $u_b=u_+$ in the bound  $(V_0>V_c)$ regime.
In the unbound regime, we solve \eqref{quartic} by treating $\omega$ as small and by setting $u=u_{p}+\delta=\delta$, where $\delta$ the first order correction to the unperturbed solution, and retaining only leading order terms in $\delta$ and $\omega$. This leads to
\begin{equation}
	u=
	%\frac{\frac{m\omega}{2\hbar}}{\sqrt{1-\frac{\pi m V_0^2}{2a_0\hbar^2\Gamma^2}}}+o(\omega)
	\frac{\frac{m_I\omega}{2\hbar}}{\sqrt{1-\left(\frac{V_0}{V_c}\right)^2}}+o(\omega)
\end{equation}
and the minimising energy at small $\omega$ reads 
\begin{equation}
	E_0(\omega)=\sqrt{1-\left(\frac{V_0}{V_c}\right)^2}\hbar \omega
\end{equation}
so that the effective mass equals 
\begin{equation}
	M^{\rm{G}}_{\rm{eff}}=\frac{m_I}{1-\left(\frac{V_0}{V_c}\right)^2}
\end{equation}
and diverges as the amplitude $V_0$ approaches the localization point. In contrast, beyond the localization point and at $\omega$ small, it is easy to see the lowest--order approximate solution of \eqref{quartic} is $b=b_++\delta$ with $\delta \sim \omega^2$, and consequently $E_0(\omega)-E_0(0)\sim \omega^2$. The limit \eqref{gaussmass} is then infinite, and one can say that the localized particle has infinite effective mass under the definition adopted here. 

Now we pass to the case $d=3$ and study the (pure) Pekar functional in $3d$
\begin{align}
	\mathcal{E}_{(2)}^{\rm{Pek}}(\psi)&=\frac{\hbar^2}{2m}\int |\nabla \psi(x)|^2 \dd x\\ \nonumber &-\frac{3}{4}\frac{\rho_0}{\ef}\iint |\psi(x_1)|^2 V^{(2)}(x_1-x_2)|\psi(x_2)|^2 \dd x_1 \dd x_2
	%\\ \nonumber&+\frac{m\omega^2}{2}\int x^2 |\psi(x)|^2dx
\end{align}
which, with our choice of the potential and wave function, evaluates to the function
\begin{equation}\label{gauss3d}
	%E_2(b)=\frac{3}{2}\frac{\hbar^2}{m_I}u-\left(\frac{m_F\rho^{1/3}V_0^2}{2\Gamma^{3/2}\hbar^2}\right)\left(\frac{u}{2u+G}\right)^{3/2}+\frac{3m_I\omega^2}{8u}.
	E_2(b)=\frac{3}{2}\frac{\hbar^2}{m_I}u-\frac{3\rho V_0^2}{4\ef}\left(\frac{\pi}{2\Gamma}\right)^{\frac{3}{2}}\left(\frac{2u}{2u+\Gamma}\right)^{\frac{3}{2}}
\end{equation}

Now, the two--body Gaussian functional \eqref{gauss3d} admits a simple calculation of the critical coupling strength for the localized state to appear as a global minimum of the untrapped system. The condition for the zero of the derivative of \eqref{gauss3d} at $\omega=0$ leads to the equation 
\begin{equation}\label{fd_z}
	\frac{(2u)^{1/2}}{(2u+\Gamma)^{5/2}}=\frac{2\hbar^2\ef}{3m_I V_0^2\rho \Gamma}\left(\frac{2\Gamma}{\pi}\right)^{\frac{3}{2}}
\end{equation}
that can be solved only if %$|V_0|\geq \left(\frac{5}{2}\right)^{5/4}\frac{\hbar^2\Gamma}{\sqrt{m_F m}}\sqrt{\frac{\Gamma^{1/2}}{\rho^{1/3}}}  $, 
$|V_0|\geq \sqrt{\frac{25\sqrt{5}}{24}\left(\frac{2}{\pi}\right)^{\frac{3}{2}}\frac{\hbar^2 \Gamma^{\frac{5}{2}}\ef}{m_I \rho}}$
which is the minimal value of $|V_0|$ for which \emph{metastable} localized states appear. The minimal value of $|V_0|$ required for \emph{stable} localization can be found by using \eqref{fd_z} together with the condition $E_2(u)=0$, as the localized and delocalized state coexist as stable minima at this value of $V_0$. This gives 
\begin{equation}
	|V_c|=\sqrt{\left(\frac{6}{\pi}\right)^{\frac{3}{2}}\frac{\hbar^2 \Gamma^{\frac{5}{2}}\ef}{2m_I \rho}}.
\end{equation} 
Unfortunately, the effective mass cannot be found explicitly, as the relevant equation for small $\omega$ is a quintic one, rather than quartic as in $2d$, and thus it does not admit a closed solution. It can be, however, computed numerically from the slope of the energy difference $E_0(\omega)-E_0(0)$ at small frequencies.

\section{Semiclassical theory of the Bose polaron}
In order to illustrate the point of our result in $d=2$, which we describe as the bosonization of the semi--classical Fermi polaron, we recall the corresponding results for the Bose polaron. We start with the Fr\"ohlich Hamiltonian describing an impurity of mass $m$ immersed in a superfluid Bose gas
\begin{equation}
	\begin{split}
		\mathbb{H}=&\int\hbar c|k| b^{\dagger}_kb_k\dd k+\frac{\hbar^2}{2m}(-\nabla_x^2)\\&+
		\int \left(\frac{|k| w_k}{\sqrt{2m_B\hbar c|k|}} b_k e^{ikx} +\mathit{h.c.}\right)dk\\&\end{split}
\end{equation}
where $c$ is the critical velocity of the gas, $w_k$ is the Fourier transform of the impurity--boson potential $W$ and $m_B$ is the mass of the bosons.  Such a Hamiltonian arises if one applies the Bogoliubov approximation to the full quantum mechanical many--body problem, and is known to be asymptotically correct in appropriate scaling regimes \cite{Tempere2009, MySe, Jonas}. Its semiclassical theory is provided by taking the expectation value of $\mathbb{H}$ on product states of the type $|\psi\rangle \otimes |z\rangle$, where $\psi$ is some impurity wave function and $|z\rangle$ is a coherent state. Minimisation over all possible coherent states yields the functional 
\begin{equation}
	\begin{split}
		\mathcal{E}_B^{\rm{Pek}}(\psi)=&\frac{\hbar^2}{2m}\int |\nabla \psi(x)|^2 \dd x\\&-\frac{1}{2m_B c^2} \iint |\psi(x)|^2 W^{(2)}(x-y)|\psi(y)|^2 \dd x \dd y
	\end{split}
\end{equation}
with $W^{(2)}=W\ast W$. As we see, the resulting semiclassical theory has the precise same structure in as we found for our model of the semiclassical Fermi polaron, provided that $d=2$. It is, in fact, known that in this particular dimension the difference between free fermions and free bosons is least prominent, for example their virial expansions are identical up to one term \cite{OlaussenSudbo}. It is known that the two systems map exactly to each other provided that suitable interactions are present \cite{NapPias}. Thus, our findings suggest that the Bose and Fermi polarons in two dimensions should display certain similarities not present in other dimensions.

\end{document}